# High quality CVD deposition of Ge layers for Ge/SiGe Quantum Well heterostructures

*Arianna Nigro[1], Eric Jutzi[1], Nicolas Forrer[1], Andrea Hofmann[1], Gerard Gadea[1,2], and Ilaria Zardo[1,2]\**

[1] University of Basel, Physics Department, Klingelbergstrasse 82, CH-4056 Basel

[2] Swiss Nanoscience Institute, Klingelbergstrasse 82, CH-4056 Basel, Switzerland

ABSTRACT. A great deal of interest is directed nowadays towards the development of innovative technologies in the field of quantum information and quantum computing, with emphasis on obtaining reliable qubits as building blocks. The realization of highly stable, controllable and accessible hole spin qubits is strongly dependent on the quality of the materials hosting them. Ultra-clean germanium/silicon-germanium heterostructures have been predicted and proven to be promising candidates and due to their large scalability potential, they are opening the door towards the development of realistic and reliable solid state, all-electric, silicon-based quantum computers. In order to obtain ultra-clean germanium/silicon-germanium heterostructures in a reverse grading approach, the understanding and control over the growth of Ge virtual substrates and thin films is key. Here, we present a detailed study on the growth kinetics, morphology, and crystal quality of Ge thin films grown via chemical vapor deposition by investigating the effects of growth temperature, partial pressure of the precursor gas and the use of Ar or $H_2$ atmosphere. The presence of carrier gases catalyzes the deposition rate and induces a smoothening on the surfaces of films grown at low temperatures. We investigated the surface roughness and threading dislocation density as a function of deposition temperature, partial pressure and gas mixture. Ge thin films deposited by diluting $GeH_4$ in Ar or $H_2$ were employed as virtual substrates for the growth of full Ge/SiGe QW heterostructures. Their defect density was analyzed and their electric transport properties were characterized via Hall measurements. Similar results were obtained for both carrier gases used.

---

\* ilaria.zardo@unibas.ch

## I. INTRODUCTION

In the last decade, group-IV materials, and in particular Germanium (Ge), have gained renewed attention due to their excellent electrical [1,2] and optical properties [3,4], which together with their compatibility with well-established CMOS manufacturing technology, allow the realization of a wide range of new devices. In particular, Germanium/Silicon-Germanium (Ge/SiGe) heterostructures are appealing platforms for photonics [5], microelectronics [6,7] and quantum computing [8]. In the frame of quantum information, they were developed as a successful planar technology able to combine low disorder [9], fast qubit driving [10] and scalability [11,12] as well as compatibility with the widespread micro-fabrication techniques employed in the integrated circuitry industry [13]. Under electrostatic gating, undoped Ge/SiGe quantum wells (QWs) accumulate holes with very high mobility, light effective mass, and tunable spin-orbit interaction allowing fast and fully electrical control of the spin qubits [14–17]. Integration over a common platform, requires a careful choice of materials. Employing Ge wafers as substrates for the heterostructure growth would be the first logical direction to follow, nonetheless, they are more expensive and less pure and durable than their silicon counterparts [18]. For these reasons, efforts were directed towards the optimization of epitaxially grown strain-relaxed Ge films on Si substrates. However, heteroepitaxial growth of Ge-rich layers on Si is challenging due to the 4.2% lattice mismatch between the two materials that induces the nucleation of dislocations [19]. Any defect in the Ge layers or at the SiGe/Ge interfaces can potentially serve as a scattering center reducing the mobility and the coherence of the spin state of the holes, compromising the qubit operation. Therefore, it is crucial to obtain single crystal, epitaxial layers and clean heterointerfaces. Various epitaxial techniques, e.g. molecular beam epitaxy (MBE) [20,21], hybrid epitaxy [22] and low energy plasma enhanced chemical vapor deposition (LEPE-CVD) [23–25] were investigated to achieve epitaxial growth of Ge-rich materials on Si. In all these cases, a forward graded $Si_{1-x}Ge_x$ alloy, in which the molar fraction x was increased from 0 to 0.7 within the alloy, was deposited on a Si substrate [2]. With this approach, several microns of material were required to minimize the nucleation of defects, resulting in elevated surface roughness and a high residual threading dislocation density, that translated into an upper bound of $10^5$ $cm^2 \cdot V^{-1} \cdot s^{-1}$ for the hole mobility in the QW region [26]. As an alternative to the forward grading, the introduction of a reverse graded buffer [27,28], grown by means of chemical vapor deposition (CVD), allowed to achieve smooth, thin heterostructures showing extremely high hole mobilities exceeding values of $10^6$ $cm^2 \cdot V^{-1} \cdot s^{-1}$ [16,29,30]. Following the reverse grading approach, a relaxed Ge film is grown on a Si substrate serving as a "virtual substrate". A linearly graded $Si_{1-x}Ge_x$ alloy, with molar fraction x varying from 1 to 0.8, is then deposited on the Ge film [28]. CVD allows the epitaxial growth of thin films with high structural quality at elevated growth rates. The growth conditions influence dramatically the growth rate, quality, and, therefore, properties of the deposited layers. Specifically, the substrate preparation, the growth temperature (affecting the reaction and surface diffusion thermally activated processes) and the partial pressures of the precursors and carrier gases (governing their impingement rate, and hence the reaction, catalysis and thermal transport kinetics) are of utmost important in CVD epitaxy processes. While this allows for a highly controllable process, the wide range of degrees of freedom makes it challenging and calls for a thorough optimization study. A systematic study of the dependence of morphology and crystal quality of the deposited layers on growth parameters is crucial for enabling the reproducibility of the experiments and their comparison.

In this work, we present a comprehensive study of the effect of the deposition conditions of Ge films on silicon substrate. Namely, we have investigated the effect of growth temperature, partial pressure of the precursor gas, and dilution in Ar or $H_2$ atmosphere on the growth kinetics, surface roughness, and defect density. Moreover, we demonstrated the performance of these Ge layers as QWs and virtual substrates via Hall effect measurements of Ge/SiGe stacks grown at selected conditions resulting from this study, and hence their applicability in highly scalable, compatible and performant spin-hole qubit planar devices.

## II. EXPERIMENTAL

All the layers within this study were grown by a cold wall CVD using a PlasmaPro 100 Nanofab reactor equipped with a showerhead (Oxford Instruments, base pressure < 0.5 mTorr), commercial Germane as a gaseous precursor ($GeH_4$, Pangas, 99.999%) and Argon and Hydrogen (Ar, $H_2$, Pangas, 99.999%) as carrier gases or Silane ($SiH_4$, Pangas 99.999%) for depositing Si-based layers in ancillary Ge/SiGe test heterostructures. Prior to growth, the Si (100) substrates (float zone, undoped, resistivity > 10'000 ohm*cm) were cleaned through a one-minute dip in 2% HF aqueous solution followed by a rinse in DI water and IPA to remove the native oxide. The investigation was performed in three steps: firstly, the growth rate of Ge thin films was studied under different conditions of precursor partial pressure, temperature, and dilution with a carrier gas. The CVD system allowed to control the flow rates and the total pressure independently – via mass flow controllers (MFCs) and an automatic valve coupled to a pressure sensor – allowing to reproducibly establish the desired partial pressure of each gas within a range of 0.1-20 Torr. Each MFC was calibrated for the specific precursor or carrier gas employed, allowing a maximum flow of 50 sccm for $GeH_4$, 2000 sccm for Ar, 2000 sccm for $H_2$, and 200 sccm for $SiH_4$. The temperature was controlled by means of a hot plate coupled to a thermocouple located under the sample, in the range of 300-600 °C. The thicknesses of the deposited Ge thin films were determined via cross-section Scanning Electron Microscopy (SEM), which together with the growth time were used to calculate the growth rates. Secondly, the morphology of the films grown with different carrier gases was characterized through Atomic Force Microscopy (AFM) in regards of surface roughness. Thirdly, full Ge/SiGe QW structures were grown on top of Ge virtual substrates obtained via Ar- and H- assisted growth. Lamellas of the QW stacks were fabricated through Focused Ion Beam (FIB) and imaged via Transmission Electron

Microscopy (TEM) to estimate the defect density in the Ge virtual substrates. The threading dislocation density (TDD) was quantitatively evaluated by means of Etch Pit Count, employing a mixture composed of HF 2.3%:HNO$_3$ 50%:CH$_3$COOH 100% in a volume ratio 10:2:2, plus 60 mg of I$_2$ dissolved per each 52 ml of mixture, and further diluted with DI water in a volume ratio 3:1. The samples were etched in the aforementioned diluted, acid and iodic mixture – adapted from [30] – for 2 minutes at room temperature, revealing etch pits capping the threading dislocations, distinguishable and quantifiable in an area basis by means of SEM. Areas of 30 x 30 μm$^2$ were defined all over the sample surface and TDD was calculated as a statistical average of the results obtained by analysing 30 images per sample. Hall-effect measurements were carried out to determine the effect of different virtual substrates on the performance of the heterostructures in terms of electronic transport properties.

III. RESULTS

A. Growth kinetics

To determine the influence of precursor partial pressure and temperature on the growth rate of Ge thin films, the cross section of the films was measured via SEM. The films were grown on 1 x 1 cm$^2$ Si chips, by using total flows of 200 sccm for 10% GeH$_4$ diluted in Ar or H$_2$ and 50 sccm for undiluted GeH$_4$. Their thickness was measured in 4 different regions, moving in steps of 2.5 mm from one side to the other, in order to account for eventual non-uniformities in the layers. Variations in thickness observed for different regions of the wafers were limited to ±2 nm for thicknesses up to 500 nm, ±5 nm for thicknesses in the range 500 nm to 2 μm, and ±10 nm for thicknesses over 2 μm. Figure 1 shows the growth rate dependence on precursor partial pressure and temperature for 10% GeH$_4$ diluted in Ar, 10% GeH$_4$ diluted in H$_2$, and undiluted

GeH$_4$. A linear increase with GeH$_4$ partial pressure and an exponential one with temperature for the growth rate can be qualitatively observed in Figure 1(a) for the three cases. The linear dependence on the precursor partial pressure can be assessed more clearly from Figure 1(b). The average slopes are 1.11 ± 0.13 mTorr min nm$^{-1}$, 1.21 ± 0.20 mTorr min nm$^{-1}$ and 0.68 ± 0.02 mTorr min nm$^{-1}$ for 10% GeH$_4$ in Ar, 10% GeH$_4$ in H$_2$, and undiluted GeH$_4$, respectively. The increase in deposition rate is explained by the higher impingement rate per unit area of molecules, which is linearly dependent on the pressure [31]. The exponential dependence on the growth temperature is highlighted through the Arrhenius plots (Figure 1(c)), which provide an activation energy for the reaction of 87.34 ± 6.78 kJ mol$^{-1}$, 80.19 ± 5.94 kJ mol$^{-1}$ and 81.33 ± 9.28 kJ mol$^{-1}$ for 10% GeH$_4$ in Ar, 10% GeH$_4$ in H$_2$ and undiluted GeH$_4$, respectively. These values of activation energy are consistent with a thermally activated process, dominated by diffusion and adsorption from the gas phase on the surface [32], and fall within the range of values reported in literature for similar reactors and process conditions (35 - 130 kJ mol$^{-1}$) [32–35]. A further trend can be identified in Figure 1(a)-(c), related to the presence and type of carrier gas mixed with GeH$_4$. Under the same conditions of reactor temperature and GeH$_4$ partial pressure, the growth rate is higher when employing a carrier gas (Ar or H$_2$) compared to the undiluted GeH$_4$ case, and a further enhancement occurs when using Ar instead of H$_2$. This behavior is explained in terms of changes in the heat transport to and from the sample upon addition of carrier gases. Adding carrier gases leads to a lower average heat capacity of the gaseous mixture [36], which leads to a decreased heat loss to the gas phase. Furthermore, the addition of a carrier gas with its own heat capacity and partial pressure contributes to the heat transfer of the gas localized in the micro-gaps between the sample and the hotplate, which due to the small scale is well into the free molecular regime [37]. Within this regime, the molecular heat flux $q_{FM}$ shows a dependence on the partial pressure of the gas mixture $P$, the thermal accommodation coefficient $\alpha$, the molar heat capacity at constant volume $C_V$, and the molecular weight $M$, by following the proportional relationship $q_{FM} \propto P \times \frac{\alpha}{2-\alpha} \times \left(C_V + \frac{R}{2}\right) \times \frac{1}{\sqrt{M}}$ [38]. The addition of a carrier gas induces an increase in the overall heat flux of the gas mixture, that translates into an improved heat transfer from the hotplate to the sample when compared to the case of undiluted $GeH_4$. By considering the values of thermal accommodation coefficients reported in literature for Ar and $H_2$, corresponding respectively to 1.1 and 0.1 evaluated accounting for the effect of multiple collisions between gas molecules [39,40], the estimated molecular heat flux is larger in the presence of Ar (3.42 W/m$^2$) compared to the case with $H_2$ (1.2 W/m$^2$). A similar calculation applied to the most common carrier gases compatible with $GeH_4$, specifically He (1.98 W/m$^2$), Ne (1.1 W/m$^2$) and $N_2$ (2.9 W/m$^2$) shows that Ar is expected to have the highest efficiency in enhancing the heat transfer. Both effects would contribute to increasing the temperature at the substrate surface, where the thermally-activated, rate-limiting $GeH_4$ dissociative adsorption occurs, leading in turn to an increase of the deposition rate. Moreover, in order to exclude differences in growth rate determined by a lower or higher velocities of gas mixtures employed, we grew Ge films by keeping the reactor temperature and the partial - and total - pressures of the precursor and carrier gases constant (500°C, $P_{GeH_4}$ = 200 mTorr, $P_{Ar/H_2}$ = 1.8 Torr), while varying the total flow rates employed. We compared the thicknesses of Ge thin films grown by using 50 sccm of 10% $GeH_4$ diluted in Ar or $H_2$, and 5 sccm of undiluted $GeH_4$, to those reported in Figure 1, for which the flow rates were set to 200 sccm of 10% $GeH_4$ diluted in Ar or $H_2$, and 50 sccm of undiluted $GeH_4$. The growth rates obtained were 18 ± 0.5 nm·min$^{-1}$ for 10% $GeH_4$ diluted in Ar, 8 ± 0.6 nm·min$^{-1}$ for 10% $GeH_4$ diluted in $H_2$ and 3 ± 0.3 nm·min$^{-1}$ for undiluted $GeH_4$ respectively, therefore in agreement with the results obtained for higher flow rates. This also confirmed the trends in growth rate as function of the carrier gases, showing that the enhancement observed when employing a carrier gas, is not determined by a higher volume or bulk speed of the gas molecules. These trends are in agreement with results previously reported in literature for similar experimental setups, that showed a higher growth rate when the Ge gas precursor was diluted in Ar, compared to the case in which it was diluted in $H_2$ [41].

In order to investigate the potential presence of a delay between the exposure to gas precursors and actual growth of the Ge films (a so-called nucleation time), the variation of the film thickness with time was evaluated for the three precursor gas mixtures at low (400°C) and high (600°C) temperature by keeping the $GeH_4$ partial pressure fixed at 500 mTorr. Figure 2 shows the film thickness dependence on the growth time. Linear fits of the data intercept the origin of the reference system, excluding any significant nucleation time.

To further investigate the effect of the dilution of $GeH_4$ in a carrier gas, we studied the variations in growth rate as function of the dilution flow rate, which is defined as the flow rate of the carrier gas (either $H_2$ or Ar) divided by the flow rate of the precursor gas (i.e. $GeH_4$), as shown in Figure 3. These experiments were performed at low (400 °C) and high (600 °C) temperature while keeping the $GeH_4$ pressure at 30 mTorr.

We observe an enhancement in growth rate with higher dilutions for both temperatures. Increasing the dilution rate while keeping the partial pressure of the precursor constant can only be achieved by increasing the carrier gas flows and the total pressure at the same time, which results in gradually increasing partial pressures of Ar and $H_2$. Increasing the flow rates and partial pressures of Ar or $H_2$ exacerbate the aforementioned heat transport effects, leading to a higher surface temperature – and hence a higher growth rate – with increasing dilution ratios. More details on how to model the dependence of the growth rate on the dilution rate can be found in the Supplemental Material S2 [42].

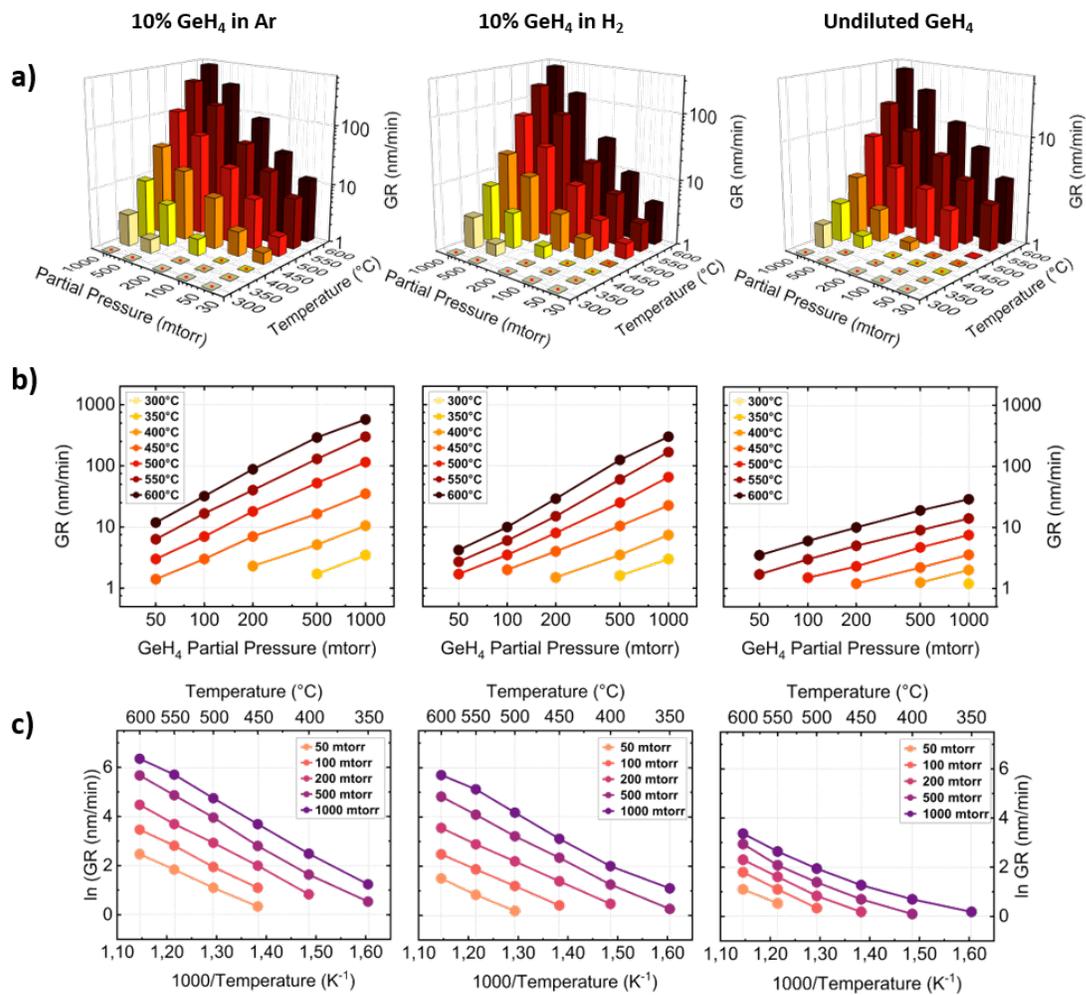

**Figure 1.** Growth rate dependence of Ge thin films on GeH$_4$ partial pressure (a)-(b) and temperature (a)-(c) for 10% GeH$_4$ in Ar, 10% GeH$_4$ in H$_2$ and undiluted GeH$_4$. Please notice that the error bars accounting for the standard deviation of the data set are plotted, but not visible due to the small values for the errors.

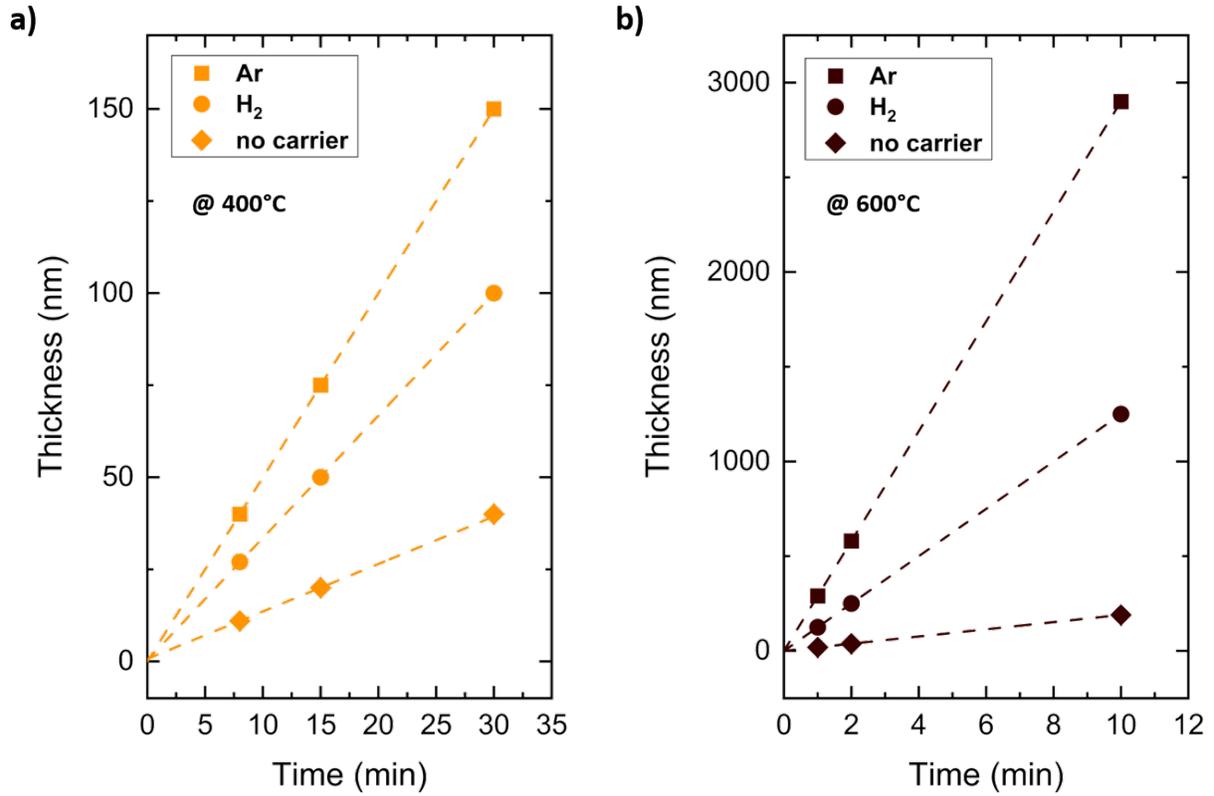

**Figure 2.** Film thickness as function of deposition time for Ge TFs deposited from 10% GeH$_4$ in Ar, 10% GeH$_4$ in H$_2$ and undiluted GeH$_4$ at (a) 400°C and (b) 600°C at a constant 500 mTorr GeH$_4$ partial pressure. Dotted lines represent a linear fit of the data sets. Please notice that the error bars accounting for the standard deviation of the data set are plotted, but not visible due to the small values for the errors.

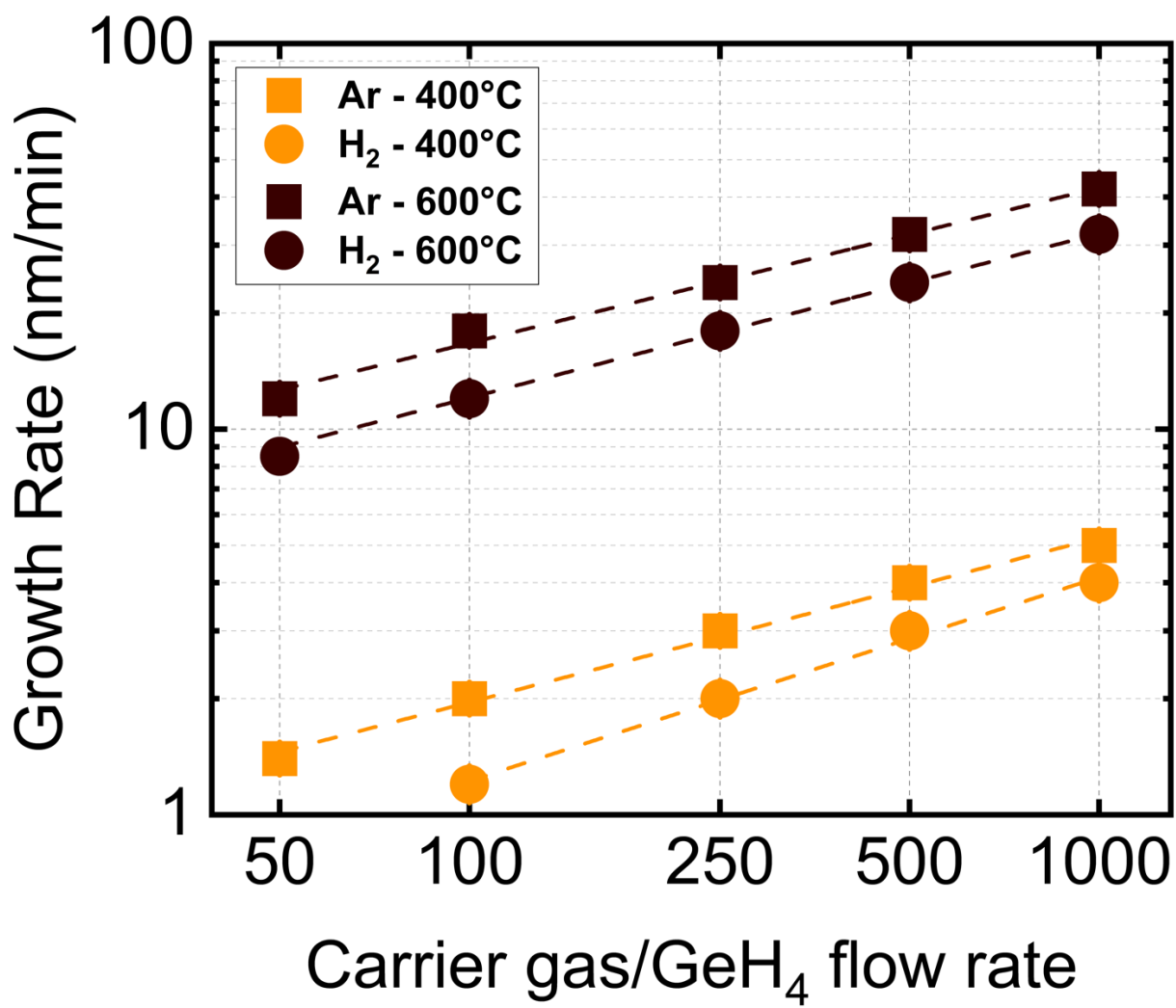

**Figure 3.** Growth rate dependence of Ge TFs on Ar and $H_2$ to $GeH_4$ flow rate at a constant 30 mTorr $GeH_4$ partial pressure at 400°C and 600°C. Dotted lines represent a linear fit of the data sets. Please notice that the error bars accounting for the standard deviation of the data set are plotted, but not visible due to the small values for the errors.

## B. Morphology

In order to study the morphology and roughness of the films and their dependence on the deposition conditions, we performed AFM measurements on 100 nm thick Ge films deposited from 10% GeH$_4$ in Ar, 10% GeH$_4$ in H$_2$ and undiluted GeH$_4$. At temperatures as low as 400°C, the films are smooth, with root mean square (RMS) roughness and maximum height difference (MHD) values below 4 nm and 35 nm, respectively, when using a carrier gas. Both roughness parameters increase when using higher temperatures, as a result of the nucleation of a larger number of Ge islands [43,44] (Figure 4(a)). No major difference in RMS and MHD is instead observed when varying the partial pressure of the precursor gas (Figure 4(b)). In both cases , the roughness of the films improves when employing a carrier gas compared to the undiluted case. To further investigate this effect, Ge thin films were grown at 400°C at a constant GeH$_4$ partial pressure of 30 mTorr and the influence of the dilution of the precursor gas on the morphology of the films was observed by measuring films deposited varying the Ar to GeH$_4$ flow rate in a range of 50 to 1000 and the H$_2$ to GeH$_4$ flow rate in a range 100 to 1000 (Figure 4 (c)-(d)). The RMS and the MHD linearly decrease when increasing the content of carrier gas, up to $1.7 \pm 0.2$ nm and $16 \pm 2$ nm for Ar and $1.3 \pm 0.1$ nm and $12 \pm 1$ nm for H$_2$, respectively, at a dilution ratio of 1000. We believe that the smoothening of the films is a consequence of a thermally induced increase of the surface diffusion of adatoms on the substrate surface, which induces the coalescence of Ge nuclei [41]. By growing a 400 nm thick additional layer at 500°C on top of the layer grown at 400 °C, the roughness and MHD decrease below 1 nm and 10 nm respectively, with values of $0.8 \pm 0.1$ nm and $1.5 \pm 0.3$ nm achieved using Ar as carrier gas and $0.7 \pm 0.2$ nm and $1.3 \pm 0.3$ nm using H$_2$ as carrier gas. Figure 5 shows the AFM 2D roughness maps for the low temperature films (panels (a) and (b)) and for the those including a high temperature layer (panels (c) and (d)) for the two cases carrier gases. The values reported herein matched well with those reported in previous works, for both low and high temperature samples [18]. Finally, we investigated the effect of the thickness and deposition temperature of the two-temperature layers on the roughness parameters. A summary of the analyzed designs is shown in Table I. For samples (a)-(c), the partial pressure of the precursor gas was set to 30 mtorr and the dilution ratio to 1000, while sample (d) was grown from undiluted $GeH_4$ at 1 torr. Also in this case, the use of a carrier gas reduces the roughness of the surface. An increase in deposition temperature for the high temperature step or a thicker low temperature layer lead to RMS values ∼ 6 nm, while increasing the thickness of the high temperature layer reduces the RMS to 3 nm. In all the cases, the measured values of RMS and MHD are considerably higher compared to those previously reported in the text for the Ar- and $H_2$- assisted films. The AFM 2D representations of the samples (a)-(d) are reported in the Supplemental Materials [42].

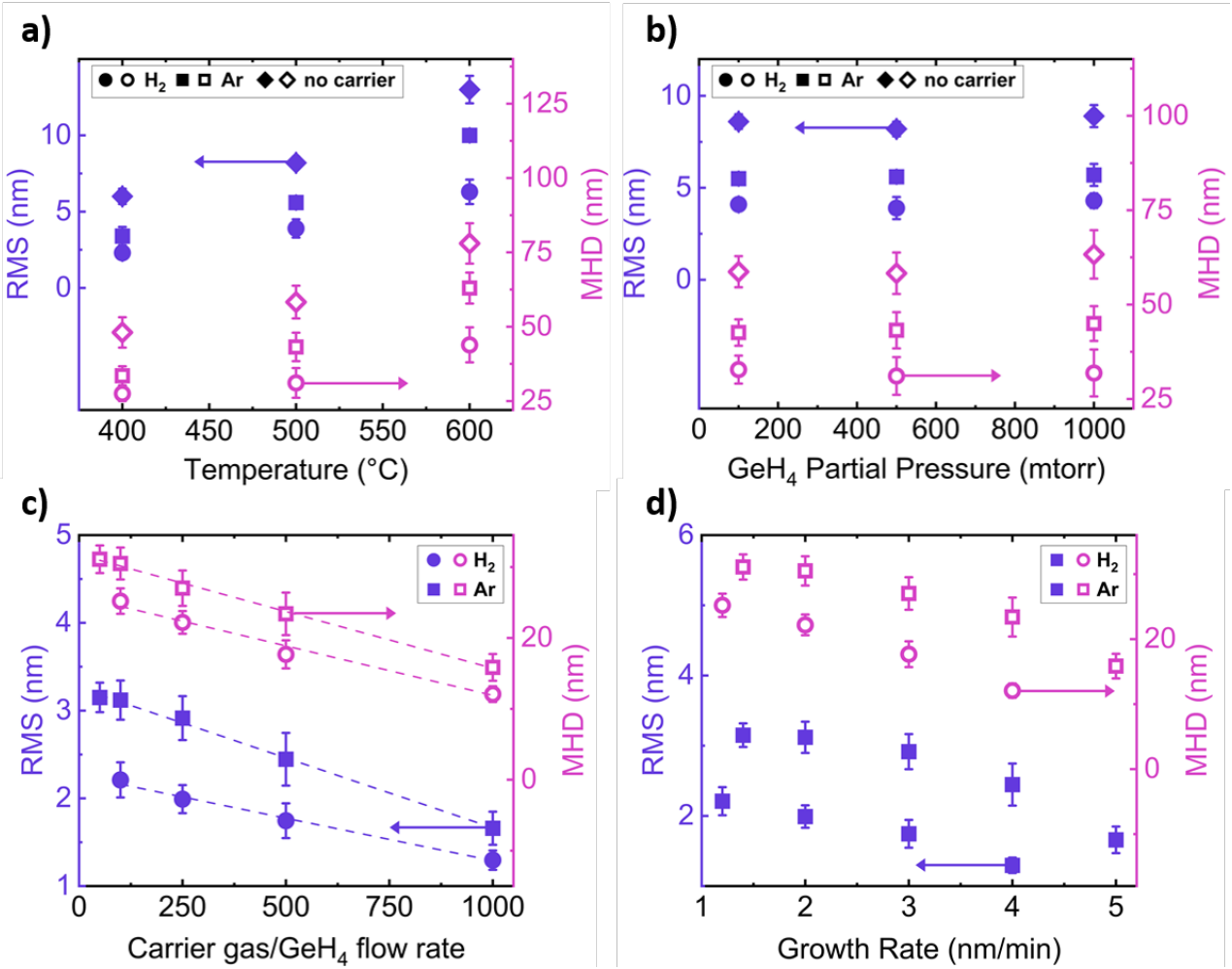

**Figure 4.** RMS roughness and MHD ranges as function of (a) temperature, for a constant GeH$_4$ partial pressure of 500 mtorr, (b) partial pressure of the precursor gas, for a constant temperature of 500°C, (c) carrier gas to GeH$_4$ flow rates and (d) corresponding growth rate for low temperature Ge thin films. Dotted lines represent a linear fit of the data sets. Please notice that axes range varies from panel to panel.

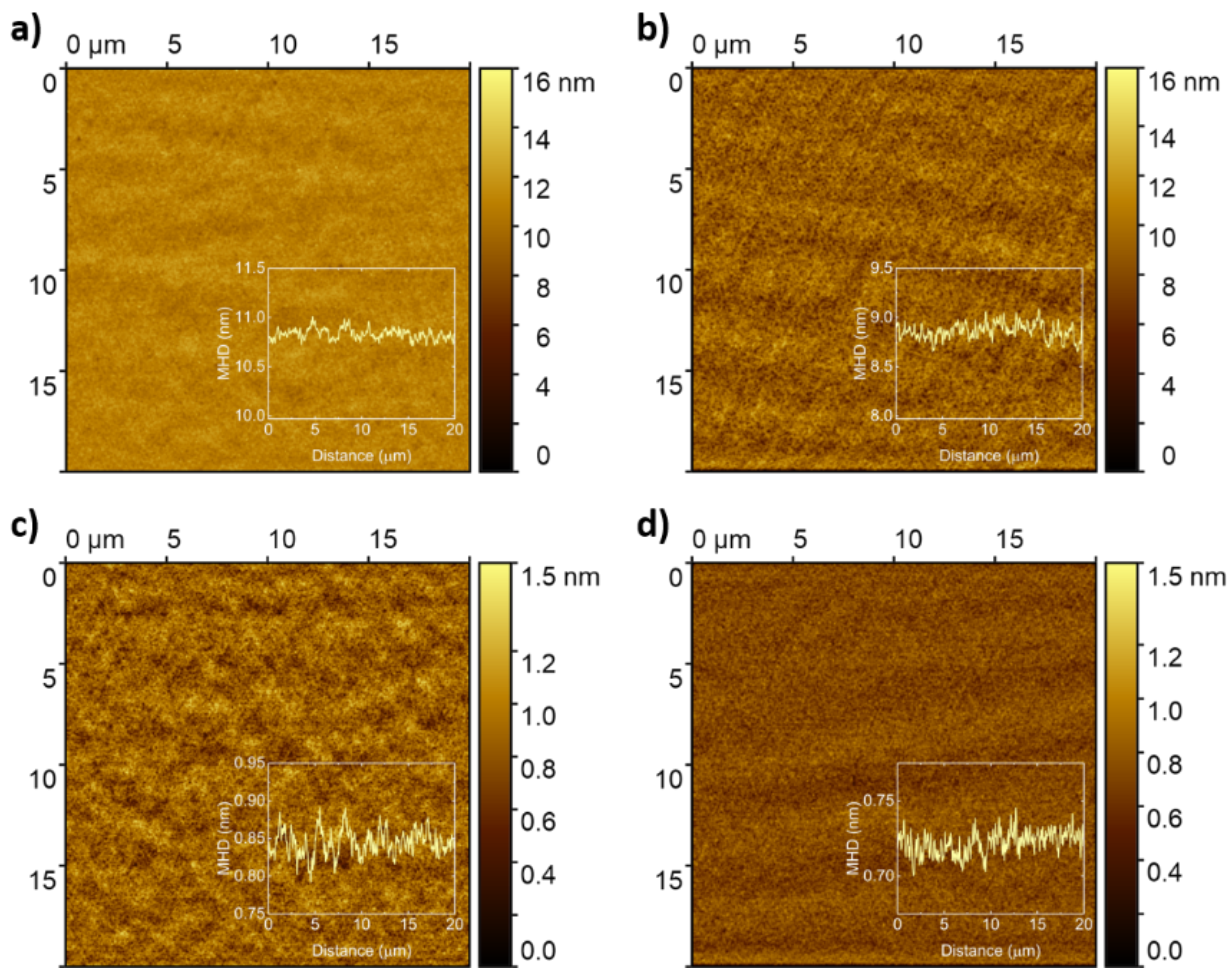

**Figure 5.** AFM 2D representation of a 100 nm low temperature film grown in (a) Ar and (b) H$_2$ atmosphere and of a 500 nm two-steps temperature Ge film grown in (c) Ar and (d) H$_2$ atmosphere. The insets to the maps show a line profile taken along the horizontal direction.

| Sample | Carrier gas | Growth T (°C) | Thickness (nm) | RMS (nm) | MHD (nm) | TDD (cm$^{-2}$) |
|---|---|---|---|---|---|---|
| (a) | Ar | 400 (LT) 600 (HT) | 100 (LT) 400 (HT) | 5.6 ± 0.8 | 54.2 ± 4.7 | (4.5 ± 0.8) x 10$^8$ |
| (b) | Ar | 400 (LT) 500 (HT) | 500 (LT) 400 (HT) | 6.2 ± 0.5 | 58.3 ± 2.6 | (4.8 ± 1.1) x 10$^9$ |
| (c) | Ar | 400 (LT) 500 (HT) | 100 (LT) 1000 (HT) | 3.1 ± 0.2 | 31.2 ± 1.7 | (4.1 ± 0.7) x 10$^9$ |
| (d) | - | 450 (LT) 550 (HT) | 100 (LT) 400 (HT) | 8.9 ± 1.1 | 65.2 ± 3.4 | - |

**TABLE I.** Overview of the layouts of the Ge virtual substrates with the respective roughness parameters and TDD obtained through AFM and EPC. The labels LT and HT stand for low and high temperature, respectively. No TDD value is given for the sample grown from undiluted GeH$_4$ due to the elevated roughness of the virtual substrate

## C. Defect density

In order to investigate the effect of the quality of a Ge virtual substrate on the transport properties of a QW heterostructure grown on top of it, we fabricated planar Ge/SiGe heterostructures through a linear reverse grading approach, as shown in Figure 6(a). We started the growth from a relaxed Ge virtual substrate ($\sim$ 500 nm), deposited on a 2 inches Si (100) substrate. The virtual substrate was grown by following the dual-step temperature approach [18], employing the deposition of a thin low temperature seed layer (400 °C, 30 mTorr $GeH_4$ partial pressure and dilution of 0.1% $GeH_4$ in $H_2$ or Ar) to induce a Frank van der Merwe growth regime and a thicker high temperature (500 °C, 400 mTorr $GeH_4$ partial pressure and dilution of 5% $GeH_4$ in $H_2$ or Ar) layer to smoothen the surface, decrease the TDD density and increase the rate of deposition of the material. The crystalline quality and the strain relaxation level were investigated by combining X-Ray diffraction (XRD), selected area electron diffraction (SAED), and Raman spectroscopy, as detailed in the Supplemental Material S3 [42]. Afterwards, we deposited a reverse linearly graded alloy ($\sim$ 750 nm, grown at 500°C, 400 to 450 mTorr $GeH_4$ partial pressure and 5% $GeH_4$ in $H_2$). The grading was achieved by keeping the $GeH_4$ and $H_2$ flow rates fixed, while increasing the $SiH_4$ flow rate, resulting into a grading rate of 10% per μm. Finally, an 80% Ge rich SiGe bottom barrier ($\sim$ 300 nm, grown at 500°C, 450 mTorr $GeH_4$ partial pressure and 5% $GeH_4$ in $H_2$), a Ge QW ($\sim$ 15 nm, grown at 500°C, 30 mTorr $GeH_4$ partial pressure and 1% $GeH_4$ in $H_2$), a $Si_{0.2}Ge_{0.8}$ top barrier ($\sim$ 55 nm, grown at 500°C, 450 mTorr $GeH_4$ partial pressure and 5% $GeH_4$ in $H_2$), and a protective Si capping layer ($\sim$ 1.5 nm, grown at 500°C and 10 Torr $SiH_4$ partial pressure) were grown to complete the heterostructure. The specific precursors' partial pressures and dilution ratio to carrier gases that were employed, were chosen in order to optimize the control over the thickness of the different layers involved..Figure 6(b) shows scanning Transmission Electron Microscopy (TEM) images of cross-sections of the obtained heterostructures in which the Ge virtual substrate was grown by diluting $GeH_4$ in Ar (left) and $H_2$ (right). The TDD can be qualitatively assessed from the images, appearing to be similar for the two cases, and by performing a selective chemical etching, namely Etch Pit Count (EPC), on the Ge virtual substrate, the TDD was estimated to be $(5.6 \pm 1.1) \times 10^9$ cm$^{-2}$ and $(5.1 \pm 1.2) \times 10^9$ cm$^{-2}$ for Ar and $H_2$ dilution, respectively. As summarized in Table I, similar values of TDD were measured when increasing the thickness of the low or high temperature layers (samples (b)-(c)), while an improvement of one order of magnitude is observed when using a higher deposition temperature for the second layer (sample (a)), as a result of dislocation annihilation through mutual interaction [19,45]. All these results are comparable to the values reported in previous works for films of the same thickness grown under similar process conditions [46], and can be expected to improve by two orders of magnitude via cyclic annealing [47]. Furthermore, we performed EPC on 100 nm thick films to quantify the TDD as function of the growth parameters for the three different gas mixtures employed. The measured densities are comparable to values previously reported in literature for thin films having the same thickness and grown with similar process conditions [18,46,48], as discussed in details in the Supplemental Material S5 [42]. Finally, Energy Dispersive X-Ray Spectroscopy (EDX) was performed to investigate the effect of Ar and $H_2$ employed in the growth of the Ge virtual substrate on the steepness of the interfaces between the Ge QW and the two SiGe barriers. The value for the steepness was extracted by fitting the chemical profile of the Si and Ge signals measured for the materials in the QW region with an error function [49,50]. Comparable values for the steepness are obtained for the QW-stack grown on top of the Ar- and $H_2$- diluted Ge virtual substrates, corresponding to 1.2 ± 0.1 nm at the interface between the bottom SiGe barrier and the Ge QW, and to 1.6 ± 0.2 nm at the interface between the Ge QW and the top SiGe barrier. These results are in good agreement with values reported in literature for heterostructures having similar layouts for the Ge QW and SiGe barriers layers [49,51].

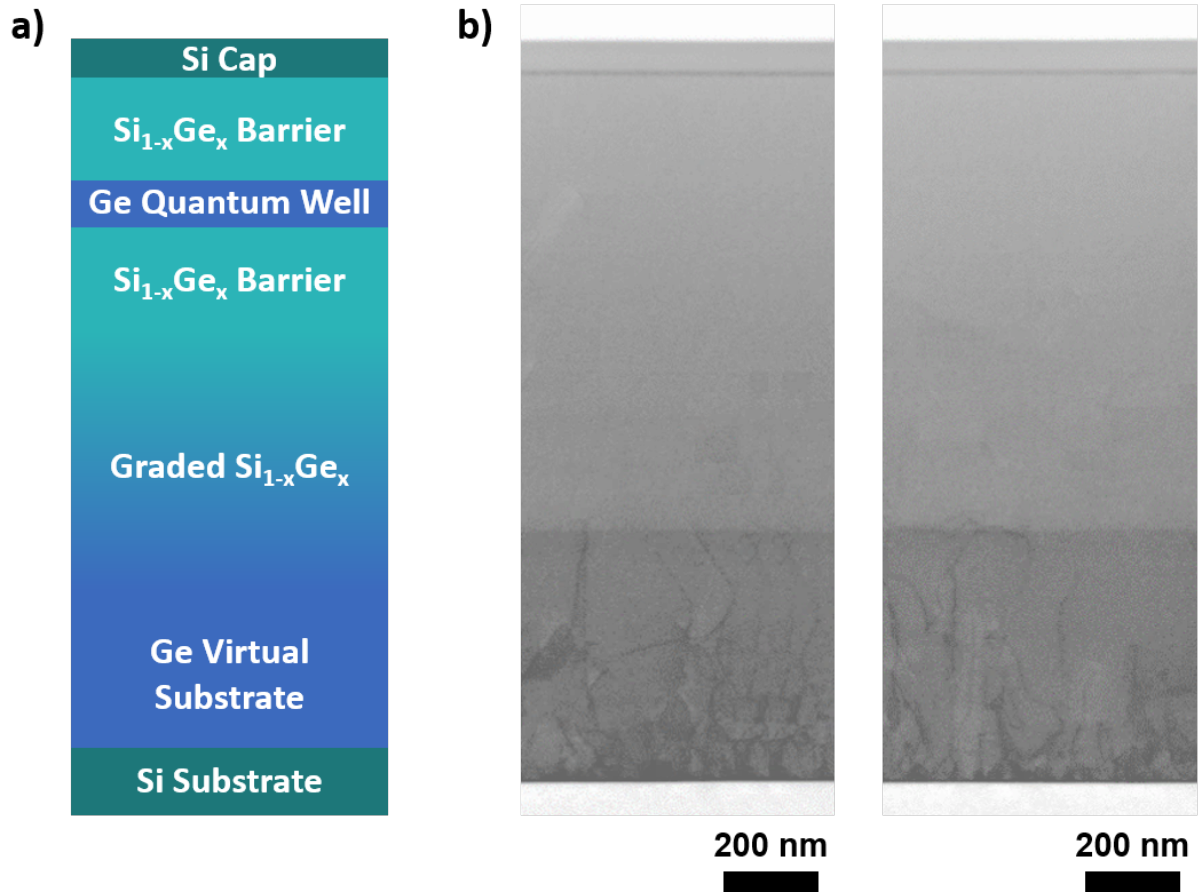

**Figure 6.** (a) Schematics of a Ge/SiGe heterostructure grown through a linear reverse grading approach. (b) STEM images of a Ge/SiGe heterostructure grown on an Ar-assisted (left) and $H_2$-assisted (right) Ge virtual substrate.

## D. Electronic transport measurements

To compare the transport behavior of QWs on top of the virtual substrate grown with Ar and the one grown with H$_2$, Hall bars as shown in Figure 7(a) of length l = 1250 μm and width w = 100 μm were fabricated simultaneously using standard UV lithography, deposition, and liftoff processes. The heterostructure was protected from bonding damage by covering the chip with a layer of Si$_3$N$_4$ after the gate lift-off and before depositing bond pads. The Si$_3$N$_4$ was grown at 300°C using a PlasmaPro 80 Plasma Enhanced CVD (PECVD) reactor (Oxford Instruments). During this step the ohmic contacts are formed by enabling the diffusion of Pt to the QW. Holes were then etched through the Si$_3$N$_4$ to contact the gate with bond pads. Both Hall bars were measured in a dilution refrigerator at a bath temperature of 7 mK. As expected from an undoped heterostructure, the QW was not populated unless holes were accumulated by applying negative DC voltages $V_g$ < -350 mV to the top-gate. A DC current of 50 nA was driven through the two-dimensional hole gas (2DHG) by applying a DC voltage bias on a 10 MΩ resistor in series with the source and drain contacts, as shown in Figure 7(a). Meanwhile, the voltages $V_{xx}$ and $V_{xy}$ were measured in order to extract the longitudinal and transversal components $\rho_{xx}$ and $\rho_{xy}$ of the resistivity tensor and the longitudinal component $\sigma_{xx}$ of the conductivity tensor. These measurements were performed at different out-of-plane magnetic field strengths B and top gate voltages $V_g$, which allows the extraction of the free carrier density $n$, the Hall mobility $\mu$, and percolation density $n_p$ [52]. At a density of 2 × 10$^{11}$ cm$^{-2}$, mobilities of 5.6 × 10$^4$ cm$^2$ V$^{-1}$ s$^{-1}$ and 4.5 × 10$^4$ cm$^2$ V$^{-1}$ s$^{-1}$ were measured for heterostructure grown with Ar- and H$_2$- assisted Ge layers, respectively. Furthermore, we estimate an effective g-factor g* = 9.1 ± 0.3 from the observed spin-splitting in the Shubnikov-de Haas oscillations [53], in agreement with [49], though large variations in g* have been reported for Ge/SiGe heterostructures [54]. We observe no beatings in the Shubnikov-de Haas oscillations and conclude that, as expected, only one sub-band is populated. Figure 7(b) shows the mobility as a function of density in both heterostructures. From fitting $\mu \sim n^\alpha$ we extracted the same exponent $\alpha = 0.70$ and $0.73$ for the heterostructures grown with Ar- and $H_2$-assisted Ge layers respectively, when comparing the two heterostructures in the same density regimes.. For the Ar assisted heterostructure, we have additional data at lower densities where $\alpha = 1.25$. The values close to $\alpha = 0.5$ indicate that the mobility is likely limited by scattering from uniform background charges in the high-density regime, while the value $\alpha = 1.25$ in the low-density regime suggests scattering from remote impurities [55] . Such impurities may reside in layers above the heterostructure, such as the gate oxide layer or its interfaces to the heterostructure or the gate metal [56].

Towards building quantum dots and for their usage in quantum computing applications, the low-energy potential landscape of the QW becomes important. The percolation density, which indicates the onset of metallic conduction in the 2DHG, may be used to benchmark this potential landscape. It is extracted by fitting the conductivity $\sigma_{xx} \sim (n-n_p)^{1.31}$ in the low-density regime [57,58]. From the data shown in Figure 7(c) we find values of $n_p = (2.39 \pm 0.02) \times 10^{10}$ cm$^{-2}$ and $(2.28 \pm 0.05) \times 10^{10}$ cm$^{-2}$ for the heterostructures grown with Ar- and $H_2$-assisted Ge layers respectively, which are comparable to previously reported numbers for Ge/SiGe heterostructures [9,29].

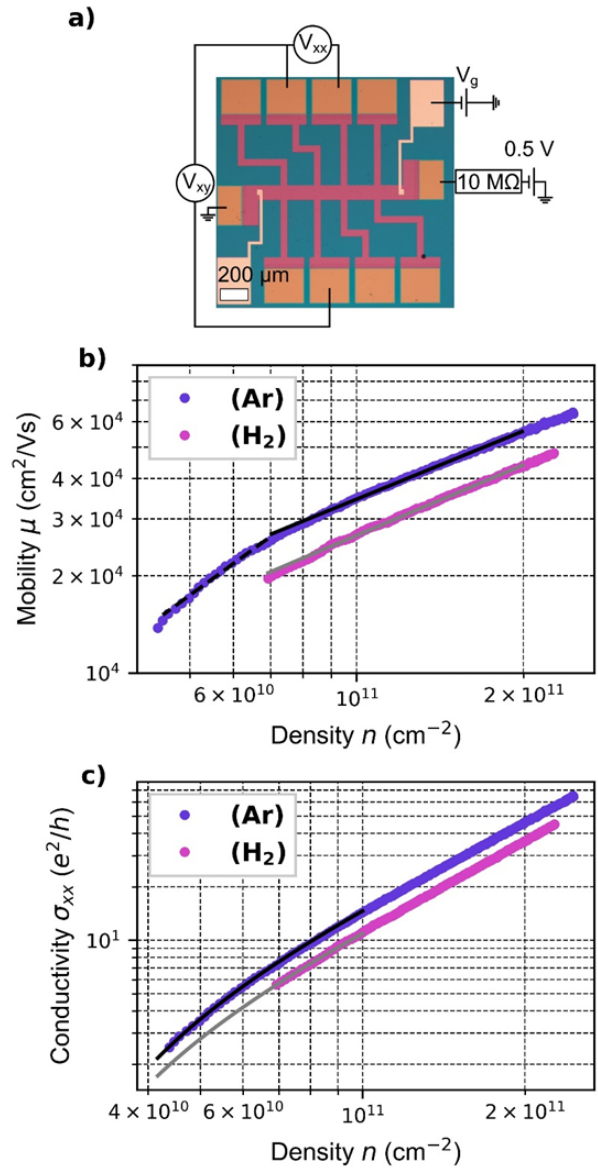

**Figure 7.** (a) Microscope image of a device nominally identical to the ones used for this study and measurement schematic. The orange squares are Pt ohmic contacts. The Ti/Au top-gate is covered by $Si_3N_4$ giving it a pink color. The two beige colored sections are Ti/Al bond pads connecting the Ti/Au gate through holes etched into the $Si_3N_4$. (b) Hall mobility as a function of density (circles) and power law fit for the common density range (solid lines) and for the low-density range (dashed lines). (c) Conductivity as a function of density (circles) and percolation density fit (lines).

## IV. DISCUSSION

Within this study, the growth kinetics of Ge thin films were studied under different conditions of temperature (300 – 600 °C), precursor partial pressure (GeH$_4$ at 50 – 1000 mTorr), and carrier gas (Ar, H$_2$ or undiluted). No growth was observed below 350°C, since the temperature is too low to overcome the activation energy for the reaction. The presence of a carrier gas is shown to catalyze the deposition rate, by thermally enhancing the dissociation rate of gas molecules, and the effect becomes more prominent when the ratio of carrier gas to GeH$_4$ is increased. No delay in deposition was observed in any of the considered cases, discarding the presence of any significant nucleation time. Low temperatures (∼ 400°C) and dilution of GeH$_4$ in Ar or H$_2$ induce the formation of monolayer islands that coalesce forming a full 2D layer typical of a Frank-van der Merwe (layer-by-layer) growth mode, keeping the surface smooth. The crystalline and surface morphological properties of the thin films were studied by Raman, XRD, SAED, EPC, and AFM measurements. All the samples are single crystalline, regardless of the deposition conditions, while films grown at high temperature exhibit a higher degree of relaxation. The TDD decreases with increasing the deposition temperature, while it is independent of GeH$_4$ partial pressure, and when using a carrier gas. As for the surface morphology, while the RMS and MHD decrease with the temperature and are independent of GeH$_4$ partial pressure, growth of high temperature Ge films onto low temperature ones ensures RMS values below 1 nm. Therefore, low temperature films are employed as seed layers when used in Ge virtual substrates, with the purpose of keeping the system planar, avoiding islanding, and generating dislocations. They are complemented by high temperatures (∼ 500°C) films to smoothen the surface and reduce the number of threading dislocations. No major differences arise in terms of roughness and defect density when using either Ar or H$_2$ as carrier gases. The percolation densities extracted from transport measurements suggest a low impurity and defect concentration, consistent with the observed smooth surfaces. The heterostructure using Ar as a carrier gas has a slightly higher mobility. As described before, no significant differences in the buffer properties are found, we know from ongoing work that small variations in the fabrication do not affect the mobility and the two samples used are similar in terms of surface and crystalline properties measured using AFM and Raman spectroscopy (more details are reported in the Supplemental Material S8 [42]). Further investigating the slight difference in mobility is out of the scope of this work.

## V.   CONCLUSIONS

A comprehensive study on the growth kinetics and morphology of Ge thin films together with electronic transport properties of Ge/SiGe heterostructures was performed for different precursor gases. Growth kinetics were characterized as a function of the growth parameters and precursor gases, showing expected linear and exponential trends with gas partial pressure and temperature respectively, and a significant enhancement when diluting GeH$_4$ in Ar or H$_2$ atmosphere. Carrier gases thermally boost the surface diffusion, hindering the formation of islands and leading to smooth layers for the layers thicknesses here considered. TDD of the Ge virtual substrates grown by two-steps temperature is in the order of $10^9$ cm$^{-2}$ regardless of the carrier gas employed. The steepness of the interfaces between the QW and the barriers and the percolation density are not affected by the choice of the carrier gas employed in the growth of the virtual substrate. Relevant improvements in the mobility values can be expected by reducing the TDD in the virtual substrate by performing cyclic thermal annealing. Finally, a further improvement of the transport and spin properties can be expected by implementing isotopically purified gas precursors.

ACKNOWLEDGMENTS

This work was supported by the Swiss National Science Foundation through NCCR SPIN (grant no. 51NF40-180604) and by the Basel QCQT PhD school. We thank Alicia Ruiz-Caridad for the TEM expertise, Nikunj Sangwan and Christian Olsen for their support in fabrication and measurements.DATA AVAILABILITY STATEMENT

The data is available at INSERT LINK

AUTHOR INFORMATION

**Author Contributions**

AN, GG, and IZ conceived the experiments. AN performed the growth studies and characterized the material with SEM, AFM, TEM and EPC. NF provided assistance for EPC and pre-growth sample preparation, EJ and AH fabricated the Hall bars, performed the transport experiments and analyzed the data. The manuscript was written through contributions of all authors. All authors have given approval to the final version of the manuscript.

ABBREVIATIONS

Ge, Germanium; Si, Silicon; SiGe, Silicon-Germanium; QW, Quantum Well; MBE, Molecular Beam Epitaxy; LEPE-CVD, Low Energy Plasma Enhanced Chemical Vapor Deposition; CVD, Chemical Vapor Deposition; SEM, Secondary Electron Microscopy; AFM, Atomic Force Microscopy; TEM, Transmission Electron Microscopy; STEM, Scanning Transmission Electron Microscopy; TDD, Threading Dislocation Density; MHD, Maximum Height Difference; RMS, Root Mean Square; EPC, Etch Pit Count; XRD, X-Ray Diffraction; SAED, Selected Area

Electron Diffraction; 2DHG, Two-Dimensional Hole Gas, Energy Dispersive X-Ray Spectroscopy (EDX).

REFERENCES

[1]   P. S. Goley and M. K. Hudait, *Germanium Based Field-Effect Transistors: Challenges and Opportunities*, Materials (Basel). **7**, 2301 (2014).

[2]   B. Rössner, D. Chrastina, G. Isella, and H. von Känel, *Scattering Mechanisms in High-Mobility Strained Ge Channels*, Appl. Phys. Lett. **84**, 3058 (2004).

[3]   C. Rödl, J. Furthmüller, J. R. Suckert, V. Armuzza, F. Bechstedt, and S. Botti, *Accurate Electronic and Optical Properties of Hexagonal Germanium for Optoelectronic Applications*, (2019).

[4]   M. Bonfanti et al., *Optical Transitions in Ge/SiGe Multiple Quantum Wells with Ge-Rich Barriers*, Phys. Rev. B (2008).

[5]   I. A. Fischer, M. Brehm, M. De Seta, G. Isella, D. J. Paul, M. Virgilio, and G. Capellini, *On-Chip Infrared Photonics with Si-Ge-Heterostructures: What Is Next?*, APL Photonics **7**, (2022).

[6]   C.-L. Chu, J.-Y. Chang, P.-Y. Chen, P.-Y. Wang, S.-H. Hsu, and D. Chou, *Ge/Si Multilayer Epitaxy and Removal of Dislocations from Ge-Nanosheet-Channel MOSFETs*, (123AD).

[7]   Y. Bogumilowicz et al., *High Ge Content Si / SiGe Heterostructures for Microelectronics and Optoelectronics Purposes*, in *Proceedings - Electrochemical Society*, Vol. 7 (2004), pp. 665–679.

[8]   G. Scappucci, C. Kloeffel, F. A. Zwanenburg, D. Loss, M. Myronov, J. J. Zhang, S. De Franceschi, G. Katsaros, and M. Veldhorst, *The Germanium Quantum Information Route*, Nat. Rev. Mater. (2020).

[9]     A. Sammak et al., *Shallow and Undoped Germanium Quantum Wells: A Playground for Spin and Hybrid Quantum Technology*, Adv. Funct. Mater. **29**, 1807613 (2019).

[10]    N. W. Hendrickx, D. P. Franke, A. Sammak, G. Scappucci, and M. Veldhorst, *Fast Two-Qubit Logic with Holes in Germanium*, Nature **577**, 487 (2020).

[11]    N. W. Hendrickx, W. I. L. Lawrie, M. Russ, F. van Riggelen, S. L. de Snoo, R. N. Schouten, A. Sammak, G. Scappucci, and M. Veldhorst, *A Four-Qubit Germanium Quantum Processor*, Nature **591**, 580 (2021).

[12]    F. Borsoi, N. W. Hendrickx, V. John, M. Meyer, S. Motz, F. van Riggelen, A. Sammak, S. L. de Snoo, G. Scappucci, and M. Veldhorst, *Shared Control of a 16 Semiconductor Quantum Dot Crossbar Array*, Nat. Nanotechnol. 1748 (2023).

[13]    G. Scappucci, P. J. Taylor, J. R. Williams, T. Ginley, and S. Law, *Crystalline Materials for Quantum Computing: Semiconductor Heterostructures and Topological Insulators Exemplars*, MRS Bull. **46**, (2021).

[14]    W. J. Hardy, C. T. Harris, Y. H. Su, Y. Chuang, J. Moussa, L. N. Maurer, J. Y. Li, T. M. Lu, and D. R. Luhman, *Single and Double Hole Quantum Dots in Strained Ge/SiGe Quantum Wells*, Nanotechnology **30**, 215202 (2019).

[15]    R. Pillarisetty, *Academic and Industry Research Progress in Germanium Nanodevices*, Nature.

[16]    Z. Kong et al., *Undoped Strained Ge Quantum Well with Ultrahigh Mobility of Two Million*, Cite This ACS Appl. Mater. Interfaces **15**, 28799 (2023).

[17]    M. Failla, M. Myronov, C. Morrison, D. R. Leadley, and J. Lloyd-Hughes, *Narrow Heavy-*

*Hole Cyclotron Resonances Split by the Cubic Rashba Spin-Orbit Interaction in Strained Germanium Quantum Wells*, Phys. Rev. B **92**, 45303 (2015).

[18] V. A. Shah, A. Dobbie, M. Myronov, and D. R. Leadley, *High Quality Relaxed Ge Layers Grown Directly on a Si(0 0 1) Substrate*, Solid. State. Electron. **62**, 189 (2011).

[19] A. E. Romanov, W. Pompe, G. Beltz, and J. S. Speck, *Modeling of Threading Dislocation Density Reduction in Heteroepitaxial Layers*, Phys. Status Solidi **198**, 599 (1996).

[20] Y. H. Xie, D. Monroe, E. A. Fitzgerald, P. J. Silverman, F. A. Thiel, and G. P. Watson;, *Very High Mobility Two-dimensional Hole Gas in Si/GexSi1−x/Ge Structures Grown by Molecular Beam Epitaxy*, Appl. Phys. Lett **16**, 2263 (1994).

[21] M. Myronov, K. Sawano, Y. Shiraki, T. Mouri, and K. M. Itoh, *Observation of Two-Dimensional Hole Gas with Mobility and Carrier Density Exceeding Those of Two-Dimensional Electron Gas at Room Temperature in the SiGe Heterostructures*, Appl. Phys. Lett. **91**, (2007).

[22] R. J. H. Morris, D. R. Leadley, R. Hammond, T. J. Grasby, T. E. Whall, and E. H. C. Parker, *Influence of Regrowth Conditions on the Hole Mobility in Strained Ge Heterostructures Produced by Hybrid Epitaxy*, J. Appl. Phys. **96**, 6470 (2004).

[23] G. Isella, D. Chrastina, B. Rössner, T. Hackbarth, H. J. Herzog, U. König, and H. von Känel, *Low-Energy Plasma-Enhanced Chemical Vapor Deposition for Strained Si and Ge Heterostructures and Devices*, Solid. State. Electron. **48**, 1317 (2004).

[24] H. Von Känel, M. Kummer, G. Isella, E. Müller, and T. Hackbarth, *Very High Hole Mobilities in Modulation-Doped Ge Quantum Wells Grown by Low-Energy Plasma*

*Enhanced Chemical Vapor Deposition*, Appl. Phys. Lett. **80**, 2922 (2002).

[25] D. Jirovec et al., *A Singlet-Triplet Hole Spin Qubit in Planar Ge*, Nat. Mater. **20**, 1106 (2021).

[26] V. A. Shah, A. Dobbie, M. Myronov, D. J. F. Fulgoni, L. J. Nash, and D. R. Leadley, *Reverse Graded Relaxed Buffers for High Ge Content SiGe Virtual Substrates*, Appl. Phys. Lett. **93**, 192103 (2008).

[27] V. A. Shah, A. Dobbie, M. Myronov, and D. R. Leadley, *Reverse Graded SiGe/Ge/Si Buffers for High-Composition Virtual Substrates*, J. Appl. Phys **107**, 64304 (2010).

[28] A. Dobbie, M. Myronov, R. J. H. Morris, A. H. A. Hassan, M. J. Prest, V. A. Shah, E. H. C. Parker, T. E. Whall, and D. R. Leadley, *Ultra-High Hole Mobility Exceeding One Million in a Strained Germanium Quantum Well*, Appl. Phys. Lett. **101**, 172108 (2012).

[29] M. Lodari, O. Kong, M. Rendell, A. Tosato, A. Sammak, M. Veldhorst, A. R. Hamilton, and G. Scappucci, *Lightly-Strained Germanium Quantum Wells with Hole Mobility Exceeding One Million*, (2021).

[30] D. Chen, Q. Guo, N. Zhang, A. Xu, B. Wang, Y. Li, and G. Wang, *High Quality Ge Epilayer on Si (1 0 0) with an Ultrathin Si1-XGex/Si Buffer Layer by RPCVD*, Mater. Res. Express **4**, 076407 (2017).

[31] R. F. C. Farrow, *Molecular Beam Epitaxy: Applications to Key Materials* (1995).

[32] B. Cunningham, J. O. Chu, and S. Akbar, *Heteroepitaxial Growth of Ge on (100) Si by Ultrahigh Vacuum, Chemical Vapor Deposition*, Appl. Phys. Lett. **59**, 3574 (1991).

[33] X. Yang and M. Tao, *A Kinetic Model for Si1−xGex Growth from SiH4 and GeH4 by CVD*, J. Electrochem. Soc. **154**, H53 (2007).

[34] M. Cao, A. Wang, and K. C. Saraswat, *Low Pressure Chemical Vapor Deposition of Si1-XGex Films on SiO2*, J. Electrochem. Soc. **142**, 1566 (1995).

[35] J.-M. Hartmann, A. Abbadie, A. M. Papon, P. Holliger, G. Rolland, T. Billon, J. M. Fédéli, M. Rouvière, L. Vivien, and S. Laval, *Reduced Pressure-Chemical Vapor Deposition of Ge Thick Layers on Si(001) for 1.3-1.55-Mm Photodetection*, J. Appl. Phys. **95**, 5905 (2004).

[36] D. C. Giancoli, *Physics Principles with Applications* (Pearson, 2014).

[37] A. Giri, P. E. Hopkins, and A. Phys Lett, *Analytical Model for Thermal Boundary Conductance and Equilibrium Thermal Accommodation Coefficient at Solid/Gas Interfaces*, J. Chem. Phys. **144**, 84705 (2016).

[38] G. S. Springer, *Heat Transfer in Rarefied Gases*, Advances in Heat Transfer.

[39] M. M. Song, S., Yovanovich, *Correlation of Thermal Accommodation Coefficient for Engineering Surfaces*.

[40] I. Yasumoto, *Accommodation Coefficients of Helium, Neon, Argon, Hydrogen, and Deuterium on Graphitized Carbon*, J. Phys. Chem. **91**, 4298 (1987).

[41] S. Kobayashi, M. Sakuraba, T. Matsuura, J. Murota, and N. Mikoshiba, *Initial Growth Characteristics of Germanium on Silicon in LPCVD Using Germane Gas*, J. Cryst. Growth **174**, 686 (1997).

[42] See Supplemental Material at LINK for additional characterizations of the crystalline and

morphological properties of the materials, and details on the transport measurements. The document includes the reference [59].

[43] P. Sheldon, B. G. Yacobi, K. M. Jones, and D. J. Dunlavy, *Growth and Characterization of GaAs/Ge Epilayers Grown on Si Substrates by Molecular Beam Epitaxy*, J. Appl. Phys. **58**, 4186 (1985).

[44] B.-Y. Tsaur, ; M W Geis, J. C. C. Fan, and ; R P Gale, *Heteroepitaxy of Vacuum-Evaporated Ge Films on Single-Crystal Si*, Appl. Phys. Lett. **38**, 779 (1981).

[45] P. Sheldon, K. M. Jones, M. M. Al-Jassim, and B. G. Yacobi, *Dislocation Density Reduction through Annihilation in Lattice-Mismatched Semiconductors Grown by Molecular-Beam Epitaxy*, J. Appl. Phys. **63**, 5609 (1988).

[46] Y. Yamamoto, P. Zaumseil, T. Arguirov, M. Kittler, and B. Tillack, *Low Threading Dislocation Density Ge Deposited on Si (1 0 0) Using RPCVD*, Solid. State. Electron. **60**, 2 (2011).

[47] Y. Yamamoto, P. Zaumseil, M. A. Schubert, and B. Tillack, *Influence of Annealing Conditions on Threading Dislocation Density in Ge Deposited on Si by Reduced Pressure Chemical Vapor Deposition*, Semicond. Sci. Technol **33**, 124007 (2018).

[48] S. Kobayashi, Y. Nishi, and K. C. Saraswat, *Effect of Isochronal Hydrogen Annealing on Surface Roughness and Threading Dislocation Density of Epitaxial Ge Films Grown on Si*, Thin Solid Films **518**, S136 (2010).

[49] A. Sammak et al., *Shallow and Undoped Germanium Quantum Wells: A Playground for Spin and Hybrid Quantum Technology - Supporting Information*, Adv. Funct. Mater.

(2019).

[50] T. E. Clark, P. Nimmatoori, K. K. Lew, L. Pan, J. M. Redwing, and E. C. Dickey, *Diameter Dependent Growth Rate and Interfacial Abruptness in Vapor-Liquid-Solid Si/Si 1-XGe x Heterostructure Nanowires*, Nano Lett. **8**, 1246 (2008).

[51] Y. Yamamoto, O. Skibitzki, M. A. Schubert, M. Scuderi, F. Reichmann, M. H. Zöllner, M. De Seta, G. Capellini, and B. Tillack, *Ge/SiGe Multiple Quantum Well Fabrication by Reduced-Pressure Chemical Vapor Deposition*, Jpn. J. Appl. Phys. **59**, SGGK10 (2020).

[52] T. Ihn, *Semiconductor Nanostructures* (New York: Oxford University Press, 2010).

[53] P. T. Coleridge, *Magnetic Field Induced Metal–Insulator Transitions in p-SiGe*, Solid State Commun. **127**, 777 (2003).

[54] M. Lodari, N. W. Hendrickx, W. I. L. Lawrie, T. K. Hsiao, L. M. K. Vandersypen, A. Sammak, M. Veldhorst, and G. Scappucci, *Low Percolation Density and Charge Noise with Holes in Germanium*, arXiv.

[55] D. Monroe, Y. H. Xie, E. A. Fitzgerald, ; P J Silverman, ; G P Watson, P. J. Silverman, and G. P. Watson, *Comparison of Mobility-Limiting Mechanisms in High-Mobility Si 1−x Ge x Heterostructures*, J. Vac. Sci. Technol. B **11**, 26 (1993).

[56] D. Laroche, S. H. Huang, Y. Chuang, J. Y. Li, C. W. Liu, and T. M. Lu, *Magneto-Transport Analysis of an Ultra-Low-Density Two-Dimensional Hole Gas in an Undoped Strained Ge/SiGe Heterostructure*, Appl. Phys. Lett. **108**, 233504 (2016).

[57] L. A. Tracy, E. H. Hwang, K. Eng, G. A. Ten Eyck, E. P. Nordberg, K. Childs, M. S. Carroll, M. P. Lilly, and S. Das Sarma, *Observation of Percolation-Induced Two-Dimensional*

*Metal-Insulator Transition in a Si MOSFET*, Phys. Rev. B - Condens. Matter Mater. Phys. **79**, 1 (2009).

[58] J. Kim, A. M. Tyryshkin, S. A. Lyon, and A. Phys Lett, *Annealing Shallow Si/SiO 2 Interface Traps in Electron-Beam Irradiated High-Mobility Metal-Oxide-Silicon Transistors*, Appl. Phys. Lett **110**, 123505 (2017).

[59] K. Oura, M. Katayama, A. V. Zotov, V. G. Lifshits, and A. A. Saranin, Growth of Thin Films, 357 (2003).